\documentclass[a4paper]{article}

\usepackage[english]{babel}
\usepackage{cmap}
\usepackage[utf8x]{inputenc}
\usepackage{amsmath, amsfonts, amsthm, amssymb}
\usepackage{graphicx}
\usepackage[colorinlistoftodos]{todonotes}
\usepackage{authblk}
\usepackage{textcomp}
\newcommand*{\No}{\textnumero}

\title{Dynamic optimization of a portfolio}

\author[1]{Oleg Malafeyev\thanks{malafeyevoa@mail.ru}}
\author[2]{Achal Awasthi\thanks{aa777@snu.edu.in}}
\affil[1]{Professor and Chair (Socioeconomic systems Modeling Department), Faculty of Applied Mathematics and Control Processes, Saint Petersburg State University, Russia}
\affil[2]{Department Of Physics, Shiv Nadar University, India}
\date{\vspace{-5ex}}
\begin{document}
\maketitle

\begin{abstract}
In this paper, we consider the problem of optimization of a portfolio consisting of securities. An investor with an initial capital, is interested in constructing a portfolio of securities. If the prices of securities change, the investor shall decide on reallocation of the portfolio. At each moment of time, the prices of securities change and the investor is interested in constructing a dynamic portfolio of securities. The investor wishes to maximize the value of his portfolio at the end of time $T$. We use a novel theoretical approach based on dynamic programming to solve the age old problem of dynamic programming. We consider two cases i.e. Deterministic and Stochastic to approach the problem and show how the portfolio is maximized using dynamic programming. 
\end{abstract}

\section{Introduction}
The problem of portfolio optimization has intrigued many people over the past few decades. Markowitz in 1952 proposed the Modern Portfolio Theory (MPT) which paved way for a new research area. The essence of Markowitz’s MPT is that there is a trade-off between risk and returns and individuals can reduce the risk associated with a single asset by diversifying their portfolios. It aims to maximize the returns of an portfolio given the risk or to minimize risk for a given value of expected returns ~\cite{MARK}.
Over the years there have been many variants to MPT. Some theories have even tried to analyze this problem by using behavioral analysis of the investors. The Behavioral Portfolio Theory (BPT) proposed by Sherfin and Statman in 2000 is a classic example of this ~\cite{STAT}. Some of the other theories like Black-Litterman Model, mean semi-variance optimization have also been proposed in the recent past. But, none of the above theories conclusively provides an answer to the ever intriguing problem of portfolio optimization. 
The problem of portfolio optimization can be considered as a dynamic process because the price of securities keep on changing at different time periods. Dynamic programming is a handy tool to analyze processes which are dynamic in nature. 
Mathematical modeling approach is widely used in different areas of science and technology. It provides deep insights if try to mathematically model various phenomenon related to economics and finance. The basics of the approaches and methodologies applied here can also be seen in [8-56]. Malafeyev and Redinskih (2014) mathematically analyze multi-agent corrupt networks using dynamic modelling. Kolokoltsov and Malafeyev in their book Understanding Game Theory (2010) and The mathematical Analysis of the Financial Markets (2010) analyzed the mathematical aspects of financial markets and have delved upon the problem of portfolio optimization. Grigorieva and Malafeev (2014), also analyze the many period postman problem with help of dynamic programming. 
In this paper, we develop a novel theoretical technique based on dynamic programming to analyze the problem of portfolio optimization.

\section{Mathematical description of the model.}
\label{dop}
Let us divide a finite interval of time $[0;T]$ into sub-intervals with the help of discrete points,
$ t_k $, $k = 1,\ldots, f\ :\ 0 = t_1 < t_2 < \ldots < t_f = T $.
At these discrete points, $l_k$ types of securities are issued, where $j_k = 1 ,\ldots, l_k $.

Let us define some important parameters related to constructing a dynamic portfolio.
\\
1) $ \tau_{j_k} $ --- is the time of maturity of securities of type $ j _k $, i.e., the interval $[t_k;t_k+\tau_{j_k}].
\mbox{ }t_k+\tau_{j_k}\leq T$, is the period during which securities of the type $j_k$ are
in circulation in the market.
\\
2) $\mu_{j_k}(t)$ --- is defined as the quotation of securities of type $j_k$ at time $t$, i.e., its market value.
\\
3) $p_{j_k}(t)$ --- is defined as the fee for the purchase and sale of securities of type $j_k$ at time $t$, for any
$j_k=1,\ldots,l_k,\mbox{ } k=1,\ldots,f$.
\\
Securities of type $j_1$ are issued at time $ t_1 $, where $j_1=1,\ldots,l_1$.
An investor decides whether or not to purchase securities of this type, at an initial state $S_0$, 

\begin{equation}
\label{dop11}
  s(t_1)=\sum_{j_1=1}^{l_1}s_{j_1}(t_1),
\end{equation}
where $s_{j_1}(t_1)$ --- is the amount which is used by the investor to buy securities type $j_1$,$j_1=1,\ldots,l_1$.
\\\\
The prices of securities of type $ j_1=1,\ldots,l_1$, at time $t_1$ are $\mu_{j_1}(t_1)$ and the investor purchases $h_{j_1}(t_1)$ number of securities of this type,
$$
h_{j_1}(t_1)=\frac{s_{j_1}(t_1)}{\mu_{j_1}(t_1)}.
$$
The value of $s(t_1)$ must satisfy the following condition:
\begin{equation}
 \label{dop12}
  s(t_1)+\sum_{j_1=1}^{l_1} p_{j_1}(t_1) \cdot h_{j_1}(t_1) \leq S_0,
 \end{equation}
where $p_{j_1}(t_1)$ is the fee paid by the investor for the purchase of securities of type $j_1$ at time $t_1$, $
j_1=1,\ldots,l_1$. The available funds that the investor has after the purchase of securities of type $j_1$, is given by the following equation:
\begin{equation}
 \label{dop13}
  S_1=S_0-s(t_1)-\sum_{j_1=1}^{l_1} p_{j_1}(t_1) \cdot h_{j_1}(t_1)
 \end{equation}
\\
At time $t_2$, the price of securities of type $j_1$, $j_1=1,\ldots,l_1$
would be -- $ \mu_{j_1}(t_2) $. So the total income of the investor at time $ t_2 $ can be defined as:
 \begin{equation}
 \label{dop14}
  W(t_2)=S_1+\sum_{j_1=1}^{l_1}\mu_{j_1}(t_2) \cdot h_{j_1}(t_1).
 \end{equation}
At time $t_2$, the price of securities of type $j_2$ also changes and this process repeats itself at each time step. For example, at time $t_3$, the price of securities of type $j_3$ changes and so on. 
\\
Now let us consider the $i$-th time period $[t_i,t_{i+1}]$.
\\
Suppose at time $t_i$ (at the beginning of the $i$-th period), an investor has $S_{i-1}$ amount of available funds . $h_{j_k}(t_{i-1})$ is defined as the number of securities of type $j_k$ which are owned by the investor at the beginning of the $i$-th period, $j_k=1,\ldots,l_k$.
\\
$W(t_i)$ is the total income of the investor at time $t_i$.
$W(t_i)$ can be mathematically expressed using the following equation:
\begin{equation}
 \label{dop15}
  W(t_i)=S_{i-1}+\sum_{k=1}^{i-1} \sum_{j_k=1}^{l_k}\mu_{j_k}(t_i) \cdot
                                 h_{j_k}(t_{i-1}).
 \end{equation}
At time $t_i$, the price of securities of type $j_i$ changes and an investor has to make a decision - whether or not to reallocate his portfolio. $s(t_i)$ is the amount which the investor has at the beginning of $i$th period. $h_{j_k}(t_{i-1})$ is the number securities that the investor owns at the beginning of the $i$-th period and  $h_{j_k}(t_i)$ is the number of securities which he wants to possess at the end of $i$-th period and at the beginning of the $(i+1)$th period. The price of securities of type $j_k=1,\ldots,l_k$,$k=1,\ldots,i$ changes in this $(t_k+\tau_{j_k} \geq t_i)$ time interval.
\\
We consider the following scenarios:
 \\
1) If $h_{j_k}(t_i)-h_{j_k}(t_{i-1})>0$, then the investor decides to re-buy these many ($h_{j_k}(t_i)-h_{j_k}(t_{i-1}$) number of securities of type $j_k$.
\\
2) If $h_{j_k}(t_i)-h_{j_k}(t_{i-1})<0$, then the investor decides to short sell these many number of ($h_{j_k}(t_i)-h_{j_k}(t_{i-1})$) securities of type $j_k$.
\\
3) If $h_{j_k}(t_i)-h_{j_k}(t_{i-1})=0$, then the investor does not change the number of securities of type $j_k$ in his portfolio in this period.
\\
4) If $h_{j_k}(t_1)=0$ and $k>1$, then there is a security of type $j_k$ at time $t_1$ whose price hasn't changed yet. Similarly, the securities, whose maturity time has already passed by $t_i$, ($t_k+\tau_{j_k}<t_i$), would be of zero value. These securities will not be of any use to the investor. 
\\
Since we have not taken inflation into account, an investor must try to invest all his funds in securities.
\\
Thus, the amount that is redistributed in the portfolio during the period $[t_i,t_{i+1}]$ is equal to:
 \begin{equation}
 \label{dop16}
  s(t_i)=\sum_{k=1}^{i} \sum_{j_k=1}^{l_k}\mu_{j_k}(t_i) \cdot
                        (h_{j_k}(t_i)-h_{j_k}(t_{i-1})).
 \end{equation}
We also have to consider the fee that brokers charge (brokerage) from their clients for the purchase and sale of the securities.$S_i$ is the available amount, which remains with the investor by the end of the period $[t_i,t_{i+1}]$. This is the amount he has at time $t_{i+1}$. $S_i$ is given by the following equation:
\begin{equation}
 \label{dop17}
  S_i=S_{i-1}-s(t_i)-\sum_{k=1}^i \sum_{j_k=1}^{l_k} p_{j_k}(t_i) \cdot
                     \mid h_{j_k}(t_i)-h_{j_k}(t_{i-1}) \mid.
  \end{equation}
The investor must select $s(t_i)$ in such a way that the value of $S_i$ be must be non-negative.
\\
The difference of $( h_{j_k}(t_i)-h_{j_k}(t_{i-1}) )$ is taken over as a small fee, which is charged during both buying and selling of securities.
\\
If there are $n$ brokerage firms in the securities market, with $r=1,\ldots,n$, then the investor chooses the firm which charges minimal commission on the deal for $j_k$-th type of securities at time $t_i$. The commission charged by brokerage firms can be mathematically expressed in the following way:
\begin{equation}
  \label{dop18}
   p_{j_k}(t_i)= \min_rp_{j_k}^r(t_i),\qquad r=1,\ldots,n,
\end{equation}
where $p_{j_k}^r(t_i)$ is the commission charged by the $r$-th brokerage firm, when facilitating a deal of security type $j_k$ in the time period $[t_i,t_{i+1}]$, ($j_k=1,\ldots,l_k$).
\\
The total income of the investor at time $T-1$ equals:
\begin{equation}
  \label{dop19}
    W(t_{T-1})=S_{T-2}+
         \sum_{k=1}^{T-2} \sum_{j_k=1}^{l_k}\mu_{j_k}
                 (t_{T-1}) \cdot h_{j_k}(t_{T-2}) .
  \end{equation}
If the investor decides to sell all existing securities at time $T-1$ then the amount by which the investor reallocates his portfolio during the period $[T-1,T]$, equals:
\begin{equation}
  \label{dop110}
     s(t_{T-1})=- \sum_{k=1}^{T-2} \sum_{j_k=1}^{l_k}\mu_{j_k}(t_{T-1})
                              \cdot h_{j_k}(t_{T-2}),
  \end{equation}
The amount which the investor has at the end of $[T-1,T]$ period equals:
  \begin{equation}
  \label{dop111}
      S_{T-1}=S_{T-2}-s(t_{T-1})-\sum_{k=1}^{T-2}
                                 \sum_{j_k=1}^{l_k}p_{j_k}(t_{T-1})
                                 \cdot h_{j_k}(t_{T-2}).
  \end{equation}
Thus, the total income of the investor at time $T$ is equal to the income of the investor at the end of the period $[T-1, T]$. This income equals:
\begin{equation}
   \label{dop112}
    W(t_T)=S_{T-1}.
   \end{equation}
The investor should try to maximize this value by carefully and dynamically changing his portfolio throughout the period $[0, T]$. This can be achieved by changing the amount allocated to different types of securities in the portfolio.

\section { Description of the problem in terms of dynamic programming.}

\subsection{\bf Deterministic case.}

We will use dynamic programming as a method of optimization.  
\\
Let us consider the case, when there is no uncertainty, i.e., all the tasks mentioned in the previous section are precisely defined.
\\
An investor decides to re-balance his portfolio at discrete time points $t_i$,  $i=1,\ldots, f-1$. This process can be divided into $f-1$ stages ( as there are $f-1$ discrete points, when the portfolio needs to be re-balanced ).
At time $t_i$, the prices of $l_i$ securities of type $j_i=1,\ldots,l_i$ are changed, and each $j_i$-th security has a maturity time, $\tau_{j_i}$. Thus, at every $i$-th step, this process can be divided into as many number of steps as the number of times the price of securities changes at time $t_i$. Thus, the intrinsic properties of the process of reallocation of funds in the portfolio allow to consider it as a multi-step process. This can be mathematically expressed as follows: 
\\

$$ (f-1)(\sum\limits_{k=1}^{f-1} \sum\limits_{j_k=1}^{l_k}l_{j_k})
$$

At each $m_i$-th step,we take those securities whose prices change at time $t_i$ and the investor needs to change his portfolio. The amount by which the portfolio changes equals:
\begin{equation}
\label{dop212}
    s_{m_i}(t_i)=\mu_{m_i}(t_i)(h_{m_i}(t_i)-h_{m_i}(t_{i-1})),
\end{equation}
\begin{equation}
\label{dop213}
    m_i=1,\ldots,\sum_{k=1}^{i}\sum_{j_k=1}^{l_k}l_{j_k}.
\end{equation}
Here, the variable $s_{m_i}(t_i)$ can be considered as a control
variable.
\\
We consider the initial amount of the investor's funds, as the initial state of the system $S_0$. 
The value $S_0, S_1,\ldots, S_{f-1}$, is the balance amount which remains with the investor, after he has reallocated his portfolio in the previous steps. We will consider these values as state parameters. 
\\
The relation (\ref{dop15}) that describes the total income of the investor for $i$ steps is the efficiency index of $i$-th step
(It is our objective function).
\\
The efficiency of the whole process is described as the income received by the investor within T periods. It can be expressed as (\ref{dop112}):
$$
W(t_T)=S_{T-1}\ ,
$$
the value of $S_{T-1}$ is given by equations
(\ref{dop110}) and (\ref{dop111}):
$$
s(t_{T-1})= -\sum_{k=1}^{T-2}\sum_{j_k=1}^{l_k}\mu_{j_k}(t_{T-1})
\cdot h_{j_k}(t_{T-2}),
$$
$$
S_{T-1}=S_{T-2}-s(t_{T-1}) -\sum_{k=1}^{T-2}\sum_{j_k=1}^{l_k}
p_{j_k}(t_{T-1})
\cdot h_{j_k}(t_{T-2}) .
$$
The state of the system described by the relations (\ref{dop16}) and
(\ref{dop17}) can be expressed as:
$$
s(t_i)= \sum_{k=1}^i\sum_{j_k=1}^{l_k} \mu_{j_k}(t_i)
\cdot(h_{j_k}(t_i)-h_{j_k}(t_{i-1})),
$$
$$
S_i=S_{i-1}-s(t_i)-\sum_{k=1}^{i}\sum_{j_k=1}^{l_k} p_{j_k}(t_i)
\cdot \mid h_{j_k}(t_i)-h_{j_k}(t_{i-1}) \mid.
$$
We introduce the function ${\cal R}(t_i,S_{i-1})$, which denotes the income received by an investor for the first $i$ steps under an optimal policy.
The function ${\cal R}(t_i,S_{i-1})$ is defined in such a way,that $i\ =\ 2,\ \ldots,\ f$ must satisfy functional equations of the form described below:
\begin{equation}
\label{dop214}
  { \cal R}(t_i,S_{i-1})=\max_{ s_{m_i}(t_i) \in D_i }
                            \{ { \cal R}(t_{i-1},S_{i-2})
                                        + \Delta W(t_i)  \},
\end{equation}
where
\begin{equation}
\label{dop215}
   \Delta W(t_i)=W(t_i)-W(t_{i-1})\ -
\end{equation}
$$ W(t_i)=S_{i-1}+\sum_{k=1}^{i-1} \sum_{j_k=1}^{l_k} \mu_{j_k}(t_i)
\cdot h_{j_k}(t_{i-1})$$
\\
How the income of an investor grows over a period $[t_{i-1},t_i]$ is subject to certain conditions.
$D_i$ is the set of admissible controls in this step, which is determined from the constraints in the problem. These controls can be mathematically expressed as follows:
$D_i$:
\begin{equation}
\label{dop216}
 0 \le \sum_{k=1}^{i} \sum_{j_k=1}^{l_k} s_{j_k}(t_i)+
             \sum_{k=1}^{i} \sum_{j_k=1}^{l_k} p_{j_k}(t_i) \cdot
             \mid h_{j_k}(t_i)-h_{j_k}(t_{i-1})\mid \le S_{i-1} .
\end{equation}
When $i=1$ we get
\begin{equation}
\label{dop217}
     { \cal R}(t_1,S_0)=S_0 \ .
\end{equation}
Equations (\ref{dop214}) and (\ref{dop217}) are the main functional equations of dynamic programming.
\\
Let us assume that at the beginning, the total income of the investor is given by:
$$
W(t_1)=S_0.
$$
Therefore, from (\ref{dop214}) and (\ref{dop217}), the function
${\cal R}(t_i)$ for $i\ =\ 2,\ldots,f$ takes the following form:
\begin{equation}
\label{dop218}
   {\cal R}(t_i)=\max_{ s_{m_i}(t_i) \in D_i }{W(t_i)}.
\end{equation}
\\
If we know the value of initial capital, we can apply the method of dynamic programming to obtain the following sequence of functions:
\\
$\{ {\cal R}(t_i,S_{i-1}) \}$ is the function corresponding to maximum income,
$ i\ =\ 2,\ldots,f$, and the corresponding vector functions
$\{(s_{m_i}^{*}(t_i))\}$ are the optimal controls at each step
$i=1,\ldots,f-1$.

\subsection{\bf Stochastic case.}

Now, let us consider a stochastic version of the deterministic process described in the previous section.The same method of using functional equations to optimize the income is also applied in the stochastic version.
\\
Let us define the prices of securities
$\mu_{j_k}(t)$, $j_k=1,\ldots,l_k$,
$k=1, \ldots, f$ $ \forall t \in [0 ; T ] $ as a discrete probability distribution $\Theta_{j_k}(t)$:
$$
\quad \mu_{j_k}(t) \quad\quad\mu_{j_k}^{(1)}(t) \quad\quad
\mu_{j_k}^{(2)}(t) \quad \quad \ldots \quad \quad\mu_{j_k}^{(n)}(t)
\quad \quad
$$
\begin{equation}
 \label{dop221}
  \Theta_{j_k}(t) \quad \quad \theta_{j_k}^{(1)}(t)\quad \quad
  \theta_{j_k}^{(2)}(t) \quad \quad \ldots \quad \quad \theta_{j_k}^{(n)}(t)
 \end{equation}
The number $n$ depends on the number of securities, i.e. $ n=n(j_k) $.
 \begin{equation}
    \label{dop222}
    \sum_{r=1}^n \theta_{j_k}^{(r)}(t)=1 ,
  \end{equation}
But $\mu_{j_k}(t_i)$ is an independent quantity. 
\\
In the stochastic case, the system goes from state $S_{i-1}$ to state $S_i$ under the influence of a vector of optimal controls
$(s_{m_i}^{*}(t_i))$ ,where $m_i$ is defined in equation (\ref{dop213}), and $ i\ =\ 1,\ldots,f-1 $. 

Since the value of $S_0$ is constant and is known to the investor in advance, the equation at time $t_1$ will be unchanged.
The functional equations of dynamic programming in this case, are as follows:
 \begin{equation}
    \label{dop223}
     \langle { \cal R }(t_i) \rangle =
                           \max_{s_{m_i}(t_i) \in D_i}
                            \{ \langle W(t_i) \rangle \} ,
   \end{equation}
   where $\langle W(t_i) \rangle$ is the mathematical expectation of the income of
investor at time $t_i$,

$\langle R(t_i) \rangle$ is the mathematical expectation of the income which is received by the investor for $i$ periods under optimal policy,
$i=2, \ldots, f$.

The prices $ \mu_{j_k}(t_i) $ of the securities are set by the probability distribution $ \Theta_{j_k}(t) $,
$ j_k =1,\ldots,l_k $, $ k=1,\ldots,f $,
$ \forall t \in [ 0,T ] $. 
The recurrence relations (\ref{dop16})--(\ref{dop18}), which describe the decision process of an investor contain a random variable. It can be expressed in the following form:
\begin{equation}
    \label{dop224}
    \langle W(t_i) \rangle =
                   \langle S_(i-1) \rangle +
                   \sum_{k=1}^{i-1}
                   \sum_{j_k=1}^{l_k}
  \left( \sum_{r=1}^{n} \theta_{j_k}^{(r)}(t_i) \mu_{j_k}^{(r)}(t_i) \right)
                                   \cdot h_{j_k}(t_{i-1}) ,
   \end{equation}
  \begin{equation}
    \label{dop225}
  \langle s(t_i) \rangle =
                 \sum_{k=1}^{i}
                 \sum_{j_k=1}^{l_k}
    \left( \sum_{r=1}^{n} \theta_{j_k}^{(r)}(t_i) \mu_{j_k}^{(r)}(t_i) \right)
    \cdot \left( h_{j_k}(t_{i})-h_{j_k}(t_{(i-1)}) \right).
   \end{equation}
It is assumed that a small brokerage fee per transaction, for sale and purchase of securities also has the same distribution as the price of securities. 
Hence (\ref{dop17}) takes the following form :
$$
\langle S_i \rangle =
\langle S_{i-1} \rangle -
\langle s(t_i) \rangle -
\sum_{k=1}^{i}
\sum_{j_k=1}^{l_k}
\left (\sum_{r=1}^{n}\theta_{j_k}^{(r)}(t_i)p_{j_k}^{(r)}(t_i)
\right)\times
$$
\begin{equation}
\label{dop226} \qquad \qquad \qquad \qquad  \times \mid
h_{j_k}(t_{i})-h_{j_k}(t_{(i-1)})
                        \mid
\end{equation}

\section{Conclusion}
It was shown that dynamic programming is a powerful and effective tool in analyzing problems related to portfolio optimization.
Also, the optimization problem in the stochastic case is to determine the set of values of control variable. The mathematical expectation of the target function is optimized on the basis of these values. This case is slightly peculiar because it is impossible to decide on the choice of control variable, if the state of the system is not known in the first step. It means that before making another decision (in the next step), we need to use the knowledge of the random values that have been observed before (in the previous steps). An investor should try to optimize his portfolio by re-balancing according to the price of the securities for maximum returns.

\end{document}